\documentclass[nofootinbib,showpacs,preprintnumbers,amsmath,amssymb,floatfix]
{revtex4}
\usepackage{epsf}

\begin{document}


\title{Description of the spin structure function $g_1$ at arbitrary $x$ and arbitrary $Q^2$
}

\vspace*{0.3 cm}

\author{B.I.~Ermolaev}
\affiliation{Ioffe Physico-Technical Institute, 194021
  St.Petersburg, Russia}
\author{M.~Greco}
\affiliation{Department of Physics and INFN, University Rome III,
Rome, Italy}
\author{S.I.~Troyan}
\affiliation{St.Petersburg Institute of Nuclear Physics, 188300
Gatchina, Russia}

\begin{abstract}
The explicit expressions describing the structure function $g_1$
at arbitrary $x$ and $Q^2$ are obtained. In the first place, they
combine the well-known DGLAP expressions for $g_1$ with the total
resummation of leading logarithms of $x$, which makes possible to
cover the kinematic region of arbitrary $x$ and large $Q^2$. In
order to cover the small-$Q^2$ region the shift $Q^2 \to Q^2 +
\mu^2$ in the large-$Q^2$ expressions for $g_1$ is suggested and
values of $\mu$ are estimated. The  expressions obtained do not
require singular factors $ x^{-a}$ in the fits for initial parton
densities.
\end{abstract}


\maketitle

\section{Introduction}
The goal of obtaining universal expressions  describing the
structure function $g_1$ at all $x$ and $Q^2$ is an attractive
task from both theoretical and phenomenological point of view.
Until recently, the only theoretical instrument to describe $g_1$
was the Standard Approach (SA) which involves the DGLAP evolution
equations\cite{dglap} and standard
 fits\cite{fits} for the initial parton densities $\delta q$ and $\delta g$.
 The fits are defined from phenomenological considerations at $x\sim
 1$ and $Q^2 = \mu^2 \sim 1$GeV$^2$.
The DGLAP equations are one-dimensional, they describe the $Q^2$
-evolution only, converting $\delta q$ and $\delta g$ into the
evolved distributions $\Delta q$ and $\Delta g$. The DGLAP
equations are theoretically grounded in the kinematical the region
\textbf{A} only:
\begin{equation}\label{rega}
\textbf{A:}~~~~~ s > Q^2 \gg \mu^2,~~~~~x \lesssim 1
\end{equation}
where we have denoted $s \equiv 2pq$, with $p$ and $q$ being the
momenta of the initial hadron and photon respectively. This leaves
the other kinematical regions uncovered. It is convenient to
specify those regions as follows:

The small-$x$ region\textbf{ B}:
\begin{equation}\label{regb}
\textbf{B:} ~~~~~ s \gg Q^2 \gg \mu^2,~~~~~x \ll 1
\end{equation}
and the small-$Q^2$ regions\textbf{ C }and\textbf{ D} :
\begin{equation}\label{regc}
\textbf{C:} ~~~~~0 \leq Q^2 \lesssim \mu^2,~~~~~x \ll 1,
\end{equation}
\begin{equation}\label{regd}
\textbf{D:} ~~~~~0 \leq Q^2 \lesssim \mu^2,~~~~~x \lesssim 1.
\end{equation}

As the matter of fact, the SA has been extended from Region
\textbf{A} to the small-$x$ Region \textbf{B}, though without any
theoretical basis. The point is that after converting $\delta q$
and $\delta g$ into $\Delta q$ and $\Delta g$ with the DGLAP
evolution equations, they should be evolved to the small-$x$
region as well. The $x$ -evolution is supposed to come from
convoluting $\Delta q$ and $\Delta g$ with the coefficient
functions $C_{DGLAP}$. However, in the leading order
$C_{DGLAP}^{LO}=1$; the NLO corrections account for one- or two-
loop contributions and neglect higher loops. This is the correct
approximation in the region \textbf{A} but becomes wrong in the
Region \textbf{B} where contributions $\sim \ln^k(1/x)$ are large
and should be accounted for to all orders in $\alpha_s$.
$C_{DGLAP}$ do no include the total resummation of the leading
logarithms of $x$ (LL), so SA requires special fits for $\delta q$
and $\delta g$. The general structure of such fits (see
Refs.~\cite{fits}) is as follows:
\begin{equation}\label{fits}
\delta q = N x^{-a} \varphi(x)
\end{equation}
where $N$ is a normalization constant; $a > 0$, so $x^{-a}$ is
singular when $x \to 0$ and $\varphi (x)$ is regular in $x$ at $x
\to 0$. In Ref.~\cite{egtfit} we showed that the role of the
factor $x^{-a}$ in Eq.~(\ref{fits}) is to mimic  the total
resummation of LL performed in Refs~\cite{egtns, egts}. Similarly
to LL, the factor $x^{-a}$ provides the steep rise to $g_1$ at
small $x$ and sets the Regge asymptotics for $g_1$ at $x \to 0$,
with the exponent $a$ being the intercept. The presence of this
factor is very important for extrapolating DGLAP into the region
\textbf{B:} When the factor $x^{-a}$ is dropped from
Eq.~(\ref{fits}), DGLAP stops to work at $x \lesssim 0.05$ (see
Ref.~\cite{egtfit} for detail). Accounting for the LL resummation
is beyond the DGLAP framework, because LL come from the phase
space not included in the DGLAP -ordering
\begin{equation}\label{order}
\mu^2  < k^2_{1~\perp} < k^2_{2~\perp}<...< Q^2~
\end{equation}
for the ladder partons ($k_{2i~\perp}$ are the transverse
components of the ladder momenta $k_i$). LL can be accounted only
when the ordering Eq.~(\ref{order}) is lifted and all
$k_{i~\perp}$ obey
\begin{equation}\label{llorder}
\mu^2 < k^2_{i~\perp} < (p+q)^2 \approx (1-x)2pq \approx 2pq~
\end{equation}
at small $x$. Replacing Eq.~(\ref{order}) by Eq.~(\ref{llorder})
leads inevitably to the change of the DGLAP parametrization
\begin{equation}\label{dglapparam}
\alpha_s^{DGLAP} = \alpha_s(Q^2)
\end{equation}
by the alternative parametrization of $\alpha_s$ given by
Eq.~(\ref{a}). This parametrization was obtained in
Ref.~\cite{egta} and was used in Refs.~\cite{egtns,egts} in order
to find explicit expressions accounting for the LL resummation for
$g_1$ in the region \textbf{B}. Obviously, those expressions
require the non-singular fits for the initial parton densities.
 Let us note that the
replacement of Eq.~(\ref{order}) by Eq.~(\ref{llorder}) brings a
more involved $\mu$ -dependence of $g_1$. Indeed,
Eq.~(\ref{order}) makes the contributions of gluon ladder rungs be
infrared (IR) stable, with $\mu$ acting as a IR cut-off for the
lowest rung and $k_{i~\perp}$ playing the role of the IR cut-off
for the
 $i+1$-rung. In contrast, Eq.~(\ref{llorder}) implies that $\mu$ acts
as the IR cut-off for every rung.

The small-$Q^2$ Regions \textbf{C} and \textbf{D} are, obviously,
beyond the reach of SA because DGLAP cannot be exploited here.
Alternatively, in Refs.~\cite{egtsmq,egthtw} we obtained
expressions for $g_1$ in the region \textbf{C} and proved that
Region C can be described through the shift $Q^2 \to Q^2 + \mu^2$
in our large-$Q^2$ formulae. Combining these results with SA
obtained in Ref.~\cite{egtfit} makes it possible to describe $g_1$
in Region\textbf{ D}. For the sake of simplicity, we present below
formulae for $g_1^{NS}$, the non-singlet component of $g_1$ only.

\section{Description of $g_1$ in the region \textbf{B}}

The total resummation of the double-logarithms (DL) and single-
logarithms of $x$ in the region \textbf{B } was done in
Refs.~\cite{egtns,egts}. In particular, the non-singlet component,
$g_1^{NS}$ of $g_1$ is
\begin{equation}
\label{gnsint} g_1^{NS}(x, Q^2) = (e^2_q/2) \int_{-\imath
\infty}^{\imath \infty} \frac{d \omega}{2\pi\imath }(1/x)^{\omega}
C_{NS}(\omega) \delta q(\omega) \exp\big( H_{NS}(\omega)
\ln(Q^2/\mu^2)\big)~,
\end{equation}
with new coefficient functions  $C_{NS}$,
\begin{equation}
\label{cns} C_{NS}(\omega) =\frac{\omega}{\omega -
H_{NS}^{(\pm)}(\omega)}
\end{equation}
and anomalous dimensions $H_{NS}$,
\begin{equation}
\label{hns} H_{NS} = (1/2) \Big[\omega - \sqrt{\omega^2 -
B(\omega)} \Big]
\end{equation}
where
\begin{equation}
\label{b} B(\omega) = (4\pi C_F (1 +  \omega/2) A(\omega) +
D(\omega))/ (2 \pi^2)~.
\end{equation}
 $ D(\omega)$ and $A(\omega)$ in Eq.~(\ref{b}) are
expressed in terms of  $\rho = \ln(1/x)$, $\eta =
\ln(\mu^2/\Lambda^2_{QCD})$, $b = (33 - 2n_f)/12\pi$ and the color
factors
 $C_F = 4/3$, $N = 3$:

\begin{equation}
\label{d} D(\omega) = \frac{2C_F}{b^2 N} \int_0^{\infty} d \rho
e^{-\omega \rho} \ln \big( \frac{\rho + \eta}{\eta}\big) \Big[
\frac{\rho + \eta}{(\rho + \eta)^2 + \pi^2} \mp
\frac{1}{\eta}\Big] ~,
\end{equation}

\begin{equation}
\label{a} A(\omega) = \frac{1}{b} \Big[\frac{\eta}{\eta^2 + \pi^2}
- \int_0^{\infty} \frac{d \rho e^{-\omega \rho}}{(\rho + \eta)^2 +
\pi^2} \Big].
\end{equation}
$H_{S}$  and $C_{NS}$ account for DL and SL contributions to all
orders in $\alpha_s$. Eqs.~(\ref{a}) and (\ref{d}) depend on the
IR cut-off $\mu$  through variable $\eta$. It is shown in
Refs.~\cite{egtns,egts} that there exists an Optimal scale for
fixing $\mu$: $\mu \approx 1$ Gev for $g_1^{NS}$  and $\mu \approx
5$
 GeV for $g_1^s$.  The arguments in favor of existence of
 the Optimal scale were
 given in Ref.~\cite{egthtw}.  Eq.~(\ref{gnsint})
 predicts that  $g_1$ exhibits the power behavior in  $x$ and $Q^2$ when $x \to 0$:
\begin{equation}
\label{gnsas}g_1^{NS} \sim \big(Q^2/x^2\big)^{\Delta_{NS}/2},~
g_1^{S} \sim \big(Q^2/x^2\big)^{\Delta_{S}/2}
\end{equation}
where the non-singlet and singlet intercepts are $\Delta_{NS} =
0.42,~\Delta_{S} = 0.86$ respectively. However the asymptotic
expressions (\ref{gnsas}) should be used  with great care:
According to Ref.~\cite{egtfit}, Eq.~(\ref{gnsas}) should not be
used at $x \gtrsim 10^{-6}$. So, Eq.~(\ref{gnsint}) should be used
instead of Eq.~(\ref{gnsas}) at available small $x$. Expressions
accounting the total resummation of LL for the singlet $g_1$ in
the region \textbf{B} were obtained in Ref.~\cite{egts}. They are
more complicated than Eq.~(\ref{gnsint}) because involve two
coefficient functions and four anomalous dimensions.

\section{Unified description of Regions A and B}

As was suggested in Ref.~\cite{egtfit}, the natural way to
describe $g_1$ in the Regions A and B is to combine the small-$x$
results with the DGLAP expressions for the coefficient functions
and anomalous dimensions of $g_1$. In particular, $g_1^{NS}$ is
again given by Eq.~(\ref{gnsint}), however with the new
coefficient function $\widetilde{C}_{NS}$ and new anomalous
dimension $\widetilde{H}_{NS}$:
\begin{eqnarray}\label{combab}
\widetilde{C}_{NS} = C_{NS} + C^{DGLAP}_{NS} - \Delta  C_{NS} \\
\nonumber \widetilde{H}_{NS} = H_{NS} + \gamma^{DGLAP}_{NS} -
\Delta H_{NS}
\end{eqnarray}
where $C_{NS}$ and $H_{NS}$ are defined in
Eqs.~(\ref{cns},\ref{hns}), $C^{DGLAP}_{NS}$ and
$\gamma^{DGLAP}_{NS}$ are the DGLAP non-singlet coefficient
function and anomalous dimension. The terms $\Delta C_{NS},~\Delta
H_{NS}$ should be introduced to avoid the double counting. In the
case when the DGLAP expressions are used in $C^{DGLAP}_{NS}$ and
$\gamma^{DGLAP}_{NS}$ with the LO accuracy,
\begin{equation}\label{deltach}
\Delta C_{NS} = 1,~~\Delta
H_{NS}=\frac{A(\omega)}{2\pi}\Big[\frac{1}{\omega}+\frac{1}{2}\Big]
\end{equation}
They are the first terms of expansions of
Eqs.~(\ref{cns},\ref{hns}) in the series in $A(\omega)$. In order
to account for the NLO terms for $C^{DGLAP}_{NS}$ and
$\gamma^{DGLAP}_{NS}$, the next terms of the expansions should be
included into $\Delta C_{NS}$ and $\Delta H_{NS}$. When
Eq.~(\ref{combab}) is substituted into Eq.~(\ref{gnsint}), we
arrive at the description of $g_1^{NS}$ covering both Regions
\textbf{ A} and\textbf{ B}. Obviously, the main contribution to
$\widetilde{C}_{NS},~\widetilde{H}_{NS}$ at Region \textbf{A}
comes from their DGLAP components. On the contrary, the total
resumation terms dominate at $x \ll 1$. When Eq.~(\ref{combab}) is
used, the initial parton densities should not include singular
factors.

\section{Description of $g_1$ in the Regions \textbf{B} and \textbf{C}}

Region \textbf{C} is defined in Eq.~(\ref{regc}). It involves
small $Q^2$, so there are no large contributions
$\ln^k(Q^2/\mu^2)$ in this region. In other words, the DGLAP
ordering of Eq.~(\ref{order}) does not make sense in the region
\textbf{C }, which makes impossible exploiting DGLAP here. In
contrast, Eq.~(\ref{order}) is not sensitive to the value of $Q^2$
and therefore the total resummation of LL does make sense in the
region \textbf{C}. In Ref.~\cite{egtsmq} we suggested that the
shift
\begin{equation}\label{shift}
Q^2  \to  Q^2 + \mu^2
\end{equation}
would allow for extrapolating our previous results (obtained in
Refs.~\cite{egtns,egts} for $g_1$ in the region \textbf{B}) into
the region \textbf{C}. Then in Ref.~\cite{egthtw} we proved this
suggestion. Therefore, applying Eq.~(\ref{shift}) to $g_1^{NS}$
leads to the following expression  for $g_1^{NS}$ valid in the
regions\textbf{ B} and\textbf{ C}:
\begin{equation}
\label{gnsbc} g_1^{NS}(x+z, Q^2) = (e^2_q/2) \int_{-\imath
\infty}^{\imath \infty} \frac{d \omega}{2\pi\imath }
\Big(\frac{1}{x+z}\Big)^{\omega} C_{NS}(\omega) \delta q(\omega)
\exp\big( H_{NS}(\omega) \ln\big((Q^2+\mu^2)/\mu^2\big)\big)~,
\end{equation}
where $z = \mu^2/2pq$. Obviously, Eq.~(\ref{gnsbc}) reproduces
Eq.~(\ref{gnsint}) in the region \textbf{B}. Expression for
$g_1^S$ looks similarly  but  more complicated, see
Refs.~\cite{egtsmq,egthtw} for detail. Let us notice that the idea
of considering DIS in the small-$Q^2$ region through the shift
Eq.~(\ref{shift}) is not new. It was introduced by Nachtmann in
Ref.~\cite{nacht}  and used after that by many authors (see e.g.
\cite{bad}), being based on different phenomenological
considerations. On the contrary, our approach is based on the
analysis of the Feynman graphs contributing to $g_1$. We also
suggest that the following values for $\mu$ should be used: for
the non-singlet component of $g_1$ $\mu = 1~$GeV and $\mu =
5.5~$GeV for the singlet $g_1$.

\section{Generalization to the Region \textbf{D}}

The generalization of the results of Sect.~IV to the Region
\textbf{D} can easily be done with replacements
\begin{equation}\label{cd}
C_{NS} \to \widetilde{C}_{NS}, ~~~~~H_{NS} \to \widetilde{H}_{NS}
\end{equation}
in Eq.~(\ref{gnsbc}), with
$\widetilde{C}_{NS},~\widetilde{H}_{NS}$ defined in
Eq.~(\ref{combab}). So, we arrive at the final result: the
 expression for $g_1$ which can be used in the Regions
\textbf{A,B,C,D} universally is
\begin{equation}\label{gnsabcd}
g_1^{NS}(x+z, Q^2) = (e^2_q/2) \int_{-\imath \infty}^{\imath
\infty} \frac{d \omega}{2\pi\imath }
\Big(\frac{1}{x+z}\Big)^{\omega} \widetilde{C}_{NS}(\omega) \delta
q(\omega) \exp\big( \widetilde{H}_{NS}(\omega)
\ln\big((Q^2+\mu^2)/\mu^2\big)\big).
\end{equation}
We remind that the expressions for the initial parton densities in
Eq.~(\ref{gnsabcd}) should not contain singular terms because the
total resummation of leading logarithms of $x$ is explicitly
included into $\widetilde{C}_{NS}$ and $\widetilde{H}_{NS}$.
\section{Prediction for the COMPASS experiments}

The COMPASS collaboration now measures the singlet $g_1^S$ at $x
\sim 10^{-3}$ and $Q^2 \lesssim 3$~GeV${^2}$, i.e. in the
kinematic region beyond the reach of DGLAP. However, our formulae
for $g_1^{NS}$ and $g_1^S$ obtained in  Refs.~\cite{egtsmq,egthtw}
cover this region. Although expressions for singlet and
non-singlet $g_1$ are different, with formulae for the singlet
being much more complicated, we can explain the essence of our
approach, using Eq.~(\ref{gnsbc}) as an illustration. According to
results of \cite{egts}, $\mu \approx 5$ GeV for $g_1^S$, so in the
COMPASS experiment $Q^2 \ll \mu^2$. It means, $\ln^k(Q^2 + \mu^2)$
can be expanded into series in $Q^2/\mu^2$, with the first term
independent of $Q^2$:
\begin{equation}\label{gssmq}
g_1^S(x+z, Q^2,\mu^2) = g_1^S(z,\mu^2) + \sum_{k=1} (Q^2/\mu^2)^k
E_k(z)
\end{equation}
where $E_k(z)$ account for the total resummation of LL of $z$ and
\begin{equation}\label{cs}
g_1^S(z,\mu^2) = (<e^2_q/2>) \int_{-\imath \infty}^{\imath \infty}
\frac{d \omega}{2\pi \imath} \big(1/z \big)^{\omega}
\big[C_{S}^q(\omega)\delta q(\omega) + C_{S}^g(\omega) \delta
g(\omega)  \big],
\end{equation}
so that $\delta q(\omega)$ and $\delta g(\omega)$ are the initial
quark and gluon densities respectively and $C_{S}^{q,g}$ are the
singlet coefficient functions. Explicit expressions for
$C_{S}^{q,g}$ are given in Refs.~\cite{egts,egtsmq}. Therefore, we
can makes the following predictions easy to be checked by COMPASS:

\subsection{Prediction 1}
In the whole COMPASS range $0 \lesssim Q^2 \lesssim 3~$GeV$^2$,
the singlet $g_1$ does not depend on $x$ regardless of the value
of $x$.
\subsection{Prediction 2}
Instead of studying experimental the $x$-dependence of $g_1^S$, it
would be much more interesting to investigate its dependence on
$2pq$ because it makes possible to estimate the ratio $\delta
g/\delta q$ (see Ref.~\cite{egtsmq} for detail).

\section{Remark on the higher twists contributions}
In the region \textbf{B} one can expand terms $\sim (Q^2 +
\mu^2)^k$ in Eq.~(\ref{gnsbc}) into series  in $(\mu^2/Q^2)^n$ and
represent $g_1^{NS}(x+z,Q^2,\mu^2)$ as follows:

\begin{equation}\label{gnshtw}
g_1^{NS}(x+z,Q^2,\mu^2)= g_1^{NS}(x,Q^2/\mu^2) + \sum_{k=1}
(\mu^2/Q^2)^k T_k~
\end{equation}
where $g_1^{NS}(x,Q^2/\mu^2)$ is given by Eq.~(\ref{gnsint});
 for explicit expressions for the factors $T_k$ see
Ref.~\cite{egthtw}. The power terms in the  rhs of
Eq.~(\ref{gnshtw}) look like the power $\sim 1/(Q^2)^k$
-corrections and therefore the lhs of Eq.~(\ref{gnshtw})  can be
interpreted as the total resummation of such corrections. These
corrections are of the perturbative origin and have nothing in
common with higher twists contributions ($\equiv HTW$). The latter
appear in the conventional analysis of experimental date on the
Polarized DIS as a discrepancy between the data and the
theoretical predictions, with $g_1^{NS}(x,Q^2/\mu^2)$ being given
by the Standard Approach:
\begin{equation}\label{dglaphtw}
g_1^{NS~exp} = g_1^{NS SA} + HTW~.
\end{equation}
Confronting Eq.~(\ref{dglaphtw}) to Eq.~(\ref{gnshtw}) leads to an
obvious conclusion: In order estimate genuine higher twists
contributions to $g_1^{NS}$, one should account, in the first
place, for the perturbative power corrections predicted by
Eq.~(\ref{gnshtw}); otherwise the estimates cannot be reliable. It
is worth mentioning that we can easily explain the empirical
observation made in the conventional analysis of experimental
data: The power corrections exist for $Q^2 > 1$ GeV$^2$ and
disappear when $Q^2 \to 1$ GeV$^2$. Indeed, in Eq.~(\ref{gnshtw})
$\mu = 1$ GeV , so the expansion in the rhs of Eq.~(\ref{gnshtw})
make sense for $Q^2 > 1$ GeV$^2$ only; at smaller $Q^2$ it should
be replaced by the expansion of Eq.~(\ref{gnsbc}) in
$(Q^2/\mu^2)^n$.
\section{Conclusion}

The extrapolation of DGLAP from the standard Region \textbf{A} to
the small-$x$ Region B involves necessarily the singular fits for
the initial parton densities without any theoretical basis. On the
contrary, the resummation of the leading logarithms of $x$ is the
straightforward and most natural way to describe $g_1$ at small
$x$. Combining this resummation with the DGLAP results leads to
the expressions for $g_1$ which can be used at large $Q^2$ and
arbitrary $x$ (Regions \textbf{A} and \textbf{B}), leaving the
initial parton densities non-singular. Then, incorporating the
shift of Eq.~(\ref{shift}) into these expressions allows us to
describe $g_1$ in the small-$Q^2$ regions (Regions \textbf{C} and
\textbf{D}) and to write down Eq.~(\ref{gnsabcd}) describing $g_1$
at the Regions \textbf{A,B,C,D.} We have used it for studying the
$g_1$ singlet at small $Q^2$ which is presently investigated by
the COMPASS collaboration. It turned out that $g_1$ in the COMPASS
kinematic region depends on $z = \mu^2/2pq$ only and practically
does not depend on $x$, even at $x \ll 1$. Numerical calculations
show that the sign of $g_1$ is positive at $z$ close to 1 and can
remain positive or become negative at smaller $z$, depending on
the ratio between $\delta g$ and $\delta q$.  To conclude, let us
notice that extrapolating DGLAP into the small-$x$ region,
although it could provide a satisfactory agreement with
experimental data, leads to various wrong statements, or
misconceptions. We enlisted the most of them in Ref.~\cite{egtep}.
Below we mention one important wrong statements not included in
Ref.~\cite{egtep}:

\textbf{Misconception:} \emph{The impact of the resummation of
leading logarithms of $x$ on the small-$x$ behavior of $g_1$ is
small.}

This statement appears when the resummation is combined with the
DGLAP expressions, similarly to Eq.~(\ref{combab}), and at the
same time the fits for the initial parton densities contain
singular factors like the one in Eq.~(\ref{fits}). Such a
procedure is inconsistent and means actually a double counting of
the logarithmic contributions: the first implicitly, through the
fits, and the second in  explicit way. It also affects the
small-$x$ asymptotics of $g_1$, leading to the incorrect values of
the intercepts of $g_1$ (see Ref.~\cite{egtfit} for more detail).

\section{Acknowledgement}
B.I.~Ermolaev is grateful to the Organizing Committee of the
workshop DSPIN-07 for financial support of his participation in
the workshop.


\begin{thebibliography}{99}

\bibitem{dglap} G.~Altarelli and G.~Parisi, Nucl.~Phys.B126 (1977) 297;
V.N.~Gribov and L.N.~Lipatov, Sov.~J.~Nucl.~Phys. 15 (1972) 438;
L.N.Lipatov, Sov.~J.~Nucl.~Phys. 20 (1972) 95; Yu.L.~Dokshitzer,
Sov.~Phys.~JETP 46 (1977) 641.

\bibitem{fits} G.~Altarelli, R.D.~Ball, S.~Forte and G.~Ridolfi,
Nucl.~Phys.~B496 (1997) 337; Acta Phys. Polon. B29(1998)1145;
E.~Leader, A.V.~Sidorov and D.B.~Stamenov. Phys. Rev. D73 (2006)
034023; J.~Blumlein, H.~Botcher. Nucl. Phys. B636 (2002) 225;
M.~Hirai at al. Phys. Rev. D69 (2004) 054021.


\bibitem{egtfit} B.I.~Ermolaev, M.~Greco and S.I.~Troyan. Phys.
Lett. B622 (2005) 93.

\bibitem{egtns} B.I.~Ermolaev, M.~Greco and S.I.~Troyan.
 Nucl.Phys.B 594 (2001)71; ibid 571(2000)137.
\bibitem{egts} B.I.~Ermolaev, M.~Greco and S.I.~Troyan. Phys.Lett.B579(321),2004.

\bibitem{egta} B.I.~Ermolaev, M.~Greco and S.I.~Troyan.  Phys.Lett.B
522(2001)57.

\bibitem{egtsmq} B.I.~Ermolaev, M.~Greco and S.I.~Troyan. Eur.Phys.J
C 50(2007)823.

\bibitem{egthtw} B.I.~Ermolaev, M.~Greco and S.I.~Troyan.
Eur.Phys.J.C51 (2007) 859.

\bibitem{egtep} B.I.~Ermolaev, M.~Greco and S.I.~Troyan.
Acta Phys. Polon. B 38 No 7 (2007) 2243 (hep-ph/0704.0341).

\bibitem{nacht} O.Nachtmann. Nucl. Phys. B 63 (1973) 237.

\bibitem{bad} B.~Badelek and J.~Kwiecinski. Z.~Phys. C 43 (1989)
251; Rew. Mod. Phys. 68, No 2 (1996)445; Phys. Lett. B 418 (1998)
229.
\end{thebibliography}
\end{document}